# The Centers of Early-Type Galaxies with HST III: Non-Parametric Recovery of Stellar Luminosity Distributions

Karl Gebhardt and Douglas Richstone

Dept. of Astronomy, Dennison Bldg., Univ. of Michigan, Ann Arbor 48109

gebhardt@astro.lsa.umich.edu,dor@astro.lsa.umich.edu

Edward A. Ajhar and Tod R. Lauer

Kitt Peak National Observatory, National Optical Astronomy Observatories, P. O. Box 26732, Tucson, AZ 85726

Yong-Ik Byun and John Kormendy

Institute for Astronomy, University of Hawaii, 2680 Woodlawn Dr., Honolulu, HI 96822

Alan Dressler

The Observatories of the Carnegie Institution of Washington, 813 Santa Barbara St., Pasadena, CA 91101

S. M. Faber and Carl Grillmair

UCO/Lick Observatories, Board of Studies in Astronomy and Astrophysics, University of California, Santa Cruz, CA 95064

and

Scott Tremaine

Canadian Institute for Theoretical Astrophysics, University of Toronto, 60 St. George St., Toronto, M5S 3H8, Canada

## ABSTRACT

We have non-parametrically determined the luminosity density profiles and their logarithmic slopes for 42 early-type galaxies observed with HST. Assuming that the isodensity contours are spheroidal, then the luminosity density is uniquely determined from the surface brightness data through the Abel equation. For nearly all the galaxies in our sample, the logarithmic slope of the luminosity density ($S = d\log\nu/d\log r$) measured at $0.1''$ (the innermost reliable measurement with the uncorrected HST) is significantly different from zero; *i.e.* most elliptical galaxies have cusps. There are only two galaxies for which an analytic core ($S \to 0$) cannot be excluded. The distribution of logarithmic slopes at $0.1''$ appears to be bimodal, confirming the conclusion of Lauer *et al.* (1995) that early-type galaxies can be divided into two types based on their surface-brightness profiles; *i.e.* those with cuspy cores and those whose steep power-law profiles continue essentially unchanged in to the resolution limit. The peaks in the slope distribution occur at $S =$ –0.8 and –1.9. More than half of the galaxies have slopes steeper than –1.0. Taken together with the recent theoretical work of Merritt & Fridman, these results suggest that many (and maybe most) elliptical galaxies are either nearly axisymmetric or spherical near the center, or slowly evolve due to the influence of stochastic orbits.

*Subject headings:*

## 1. Introduction

The mass density distribution for galaxies has important consequences for the phase space structure of the galaxy (Gerhard 1987, Hasan & Norman 1990, Lees & Schwarzschild 1992, Merritt & Fridman 1995, 1996). In particular, Merritt & Fridman (1996) studied the influence of a central density cusp on the orbital structure by following orbits in a potential with a cusp and building self-consistent models from the orbit library. They have shown that in a triaxial potential with a density varying as $\rho \propto r^{-2}$, which they term a "strong cusp", a fraction 80% of the orbits are irregular, that is, they conserve energy but have no other isolating integrals. This behavior is caused by scattering off the steep inner potential spike. They also showed that if $\rho \propto r^{-1}$, which they term a "weak cusp", a fraction 60%–80% (depending on energy) of the orbits are irregular. This is in sharp contrast with separable triaxial potentials, in which all of the orbits are regular (reviewed in de Zeeuw & Franx 1991). In all of these potentials, the density is well behaved (Taylor series expandable) near the center and, therefore, they can be described as having "analytic" cores.

Merritt and Fridman's work strongly suggests that in general, if the density distribution has a nonzero logarithmic slope near the center, the orbits that pass near the center (*i.e.*, box orbits) will be chaotic. Therefore, if the density distribution is triaxial and has $d \log \rho / d \log r \sim -1/2$, or steeper, then stochasticity is likely to play a critical role in the structure and evolution of the central regions of the galaxy (Merritt & Fridman 1995). The role of a central cusp is less important in the axisymmetric case since orbits do not pass near the center because they conserve one component of the angular momentum. Merritt & Fridman's work emphasizes the important role of high resolution surface brightness distributions in understanding the dynamics of elliptical galaxies and spiral bulges.

Lauer *et al.* (1995, Paper 1) have presented surface brightness profiles for early-type galaxies observed with the uncorrected HST. These can be deprojected to obtain the luminosity density which we assume also represents the mass density profile. The existence of a central massive dark object (MDO) will only increase the central mass concentration, so the luminosity density probably provides a lower limit for the mass concentration. Deprojection to obtain the mass density is formally simple in the case of spherical (or spheroidal) symmetry. If the galaxy has a luminosity density $\nu(r)$, then the surface brightness as a function of projected radius is

$$I(R) = \int_R^\infty \frac{\nu(r)\, r\, dr}{\sqrt{r^2 - R^2}}. \tag{1}$$

This equation is an Abel integral equation with solution given by

$$\nu(r) = -\frac{1}{\pi} \int_r^\infty \frac{d\,I}{dR} \frac{dR}{\sqrt{R^2 - r^2}}\ . \tag{2}$$

Cast in this way, the solution looks deceptively simple. In practice any noise in the data is amplified by the construction of $\nu(r)$, and further amplified by a second differentiation to construct $S \equiv d\log\nu(r)/d\log r$. In Paper 1, we followed the time honored procedure of fitting a parametric model to the data before deprojecting. The model had the form

$$I(R) = \frac{2^{\beta-\alpha} I_b}{(r/r_b)^\gamma [1 + (r/r_b)^\alpha]^{(\beta-\gamma)/\alpha}} \tag{3}$$

This parametric model was fitted to the observed data by a maximum-likelihood technique (Byun *et al.* 1996, Paper II). The parametric model can then be inverted numerically using Eqn. 2 to estimate the stellar density run near the center. While this approach is reasonable and should yield good estimates of $\nu$ if the parametric model fits the data well, there is some cause for concern in its application.

It is easiest to understand this concern by contemplating Figure 3 and eqn 28 of Tremaine *et al.* (1994) (see also Dehnen 1993). That paper examined a variety of properties of "$\eta-$ models", with *density* profiles of the form

$$\rho_\eta(r) = \frac{\eta}{4\pi} \frac{1}{r^{3-\eta}(1+r)^{1+\eta}}. \tag{4}$$

The figure and eqn 28 show that for $2 < \eta < 3$ the central surface brightness is finite (and therefore $d\log I(R)/d\log R = 0$), despite the fact that the central density is infinite and the central value of $d\log\nu(r)/d\log r$ varies, in this subset of those models, over the open interval $(-1, 0)$. The problem is that in a model with a mild central density cusp, the surface brightness (see equation 1) near $R \sim 0$ is not dominated by the density near $r \sim 0$. Rather it is the behavior of $\nu(r)$ near the break radius ($r \sim 1$ in the $\eta-$models), that determines the surface brightness near zero projected radius. An alternative and more frightening way to put this is that the use of Eqn 3 as a parametric model creates a *discontinuous* mapping from the surface brightness to the luminosity density near $\gamma = 0$. If the model has $\gamma$ exactly $= 0$ and $\alpha = 2$, then the deprojected model will have an analytic core. For any value of $\gamma$ greater than zero, even by an arbitrarily small amount, deprojecting either Eqn. 3 or 4 will lead to an asymptotic value of $d\log\nu(r)/d\log r$ *at least* as steep as –1. However, the radii at which this discontinuity occurs are smaller than the HST resolution for most of the galaxies in our sample, and this effect will not be as significant. This point is also illustrated in Figure 8 in Paper 1.

This argument suggests that estimates of the density and its derivative at small radii may be suspect in galaxies with relatively flat inner surface brightness profiles. In view of the importance of the density slope in determining dynamics and evolution of the center of the galaxy, and because of the potential sensitivity of some of our earlier conclusions to this problem, we have recomputed the density and its derivative at $0.1''$ using non-parametric methods in all of the galaxies reported in Paper 1 and some additional galaxies (a similar analysis was done on a smaller sample of six galaxies by Merritt & Fridman 1995). Various methods to handle equation 2 have been discussed in the statistics literature and many of these are now being introduced to astronomy (Merritt & Tremblay 1994, Gebhardt & Fischer 1995). In the present case, two obvious choices are a kernel estimator or a smoothing spline fit to the data. We choose the latter method and describe it in some detail below.

We also rediscuss the question of whether there appear to be two classes of objects in this sample, based on the logarithmic slope of the central density. The existence of two classes continues to appear probable.

## 2. Data

The data comprise the surface brightness profiles for the 45 galaxies in Paper 1 and 21 galaxies taken from the HST archive (presented in Byun *et al.* 1996). All of the measurements are from the uncorrected HST. The 21 additional galaxies, originally observed as part of various GO and GTO programs, were obtained from the HST archive when they became publicly available. All images were deconvolved using 80 iterations of the the Lucy-Richardson algorithm (Lucy 1974; Richardson 1972) and the same, high signal-to-noise composite PSFs used in Paper 1. As in Paper 1, we used the PSF closest in time to the observation date for each galaxy. Measurement of surface brightness profiles was described in detail in Paper 1. The central density slope is not reported for those galaxies which have an AGN, inner dust obscuration, or central flat-fielding defects. This was 35% of the total initial sample, and we therefore have 42 remaining galaxies which are used in the following analysis. Included in this sample are 12 galaxies which show some dusty areas in the central regions, but we were adequately able to exclude those areas when calculating the surface brightness profiles. The results do not change if those galaxies are excluded from the sample.

The HST profiles extend out to $10''$. Formally, the solution of Eqn. 2 requires $I(R)$ at all $R$. For $R > 10''$ we have used a de Vaucouleurs profile. The effective radius was either taken from Faber *et al.* (1989), or from the Second Reference Catalogue of Bright Galaxies (de Vaucouleurs *et al.* 1976). If the effective radius was not available then it was approximated based on the best fitted de Vaucouleurs profile to the outer regions of the HST data. Errors in the surface brightness at large radii have a negligible effect on the central luminosity density and the estimated central slopes.

## 3. Non-Parametric Deprojection

As noted in the introduction, inverting the projected light distribution of a symmetrical object is an ill-conditioned problem that has gotten considerable attention from statisticians. A number of possible techniques can be applied, which use either density estimation or regression analysis. Merritt & Tremblay (1994) used density estimation methods. These use a kernel representation for the surface brightness data to provide a smooth probability density, from which the luminosity density is directly inferred. Here, we will use regression analysis (as in Merritt 1993, Gebhardt & Fischer 1995, and Merritt & Fridman 1995). Regression techniques involve fitting a smooth curve to the data points, using either kernel estimators (Scott 1992, p. 229) or as we do here using penalized likelihood. In our case, a smooth curve is fit to $dI/dR$, and $\nu(r)$ is obtained through numerical integration of Eqn. 2. Both techniques have their relative merits; density estimation *directly* provides the luminosity density, whereas regression analysis requires smoothing the surface brightness but it is computationally simple and efficient.

The HST data provide a very smooth projected density, whose accuracy is limited primarily by dust, systematic errors of our ellipse fitting method, and calibration of the WFPC rather than by Poisson noise. Lauer *et al.* (1995) and Kormendy *et al.* (1995) have separately modelled the effects of miscalibrating the HST PSF and conclude that they will not affect our results. Since smoothing techniques are designed to overcome noise in the data and since the dominant source of error in the surface brightness is not Poisson noise, we believe that our choice of the specific smoothing method will have little impact on the results.

We have chosen to use smoothing splines to represent the surface brightness profiles (as a function of R) as derived from the processing of the raw data in Paper 1. The choice of splines for this purpose is motivated by their dual qualities as statistical estimators and functional approximators, as discussed by Wabha (1990). We construct the smoothing spline by tabulating surface brightness versus projected radius on a log-log scale. For a given stiffness, $\lambda$, the best smoothing spline, $g$, minimizes

$$F_\lambda = \sum_{i=1}^{N} \frac{(y_i - g(x_i))^2}{\sigma_i^2} + \lambda \int (g''(x))^2 dx, \qquad (5)$$

where $y_i$ are the data, $\sigma_i$ are the data uncertainties and $x = \log r$. The second term in the above equation is the penalty function. In the limit of infinite smoothing ($\lambda \to \infty$), the spline is forced to a straight line ($g'' \to 0$). In our case, since $y_i$ is $\propto \log I$ and $x = \log R$, a straight line fit is a power law. This is due to our choice of the second derivative in the penalty function. A wise choice of $\lambda$ trades fidelity to the data against the smoothness of the spline fit, and one should make the choice so as to achieve an acceptable value of $\chi^2$ in the first term in equation 5. Unfortunately, the number of degrees of freedom in this problem is not easily defined. In the limit $\lambda \to 0$, there are $N$ data points, $N$ spline equations and $2N - 2$ unknown spline coefficients (the values and second derivatives of the spline fit at the $x_i$). Clearly, the number of degrees of freedom in this case is $\sim 0$, and the spline fit goes through each data point with $\chi^2 = 0$. If, on the other hand, we choose $\lambda$ very large, the spline second derivatives are driven toward 0, and the number of degrees of freedom tends toward $N - 2$, which is reasonable since we are fitting a straight line to the data (in our case, a power law).

In situations where the uncertainties are known, statisticians advocate the use of the "predicted mean error" to estimate the optimal smoothing. One can show (Craven and Wahba 1979) that the appropriate choice of $\lambda$ in this case minimizes, subject to equation 5, the quantity $R$ given by

$$R = N^{-1} \left( ||(I - A)y||^2 - \sigma^2 \left( \text{Trace } [(I - A)^2] - \text{Trace } (A^2) \right) \right). \qquad (6)$$

where $A$ is the spline coefficient matrix such that $g = Ay$ (i.e., $A$ contains the location and second derivatives for the spline estimate), $I$ is the identity matrix, $\sigma$ is the average value of the known data uncertainties, and $||a||$ is the norm of vector $a$. Eqn. 6 is derived from the sum of the squared differences between the estimated and actual underlying function.

Alternatively, if the data uncertainties are unknown or are suspect, one can use generalized cross validation (GCV) to estimate the optimal smoothing. GCV uses a jackknife approach. For a given $\lambda$, one point is removed from the sample and the spline from the remaining $N - 1$ points is used to estimate the value for the removed point. This is repeated for each point, and the optimal smoothing is that which minimizes the sum of squared differences of the actual data points from the estimated points. A good explanation of GCV can be found in Craven & Wahba (1979) and Wahba (1990). GCV also provides an estimate of the average error variance of the data points. For the present data, the error variance as determined by GCV implies approximately 0.01 average uncertainty ($1\sigma$) in the log surface density. This is similar to the quoted value of Lauer *et al.* (1995) and Byun *et al.* (1996). Therefore, using either "predicted mean error" with 0.01 uncertainty or GCV gives the same result. We use "predicted mean error" for all galaxies since it is always better to use the data uncertainties, if known, when choosing the optimal smoothing (Craven & Wahba 1979).

The smoothing value, as explained above, is chosen to be optimal for the surface brightness data. However we are interested in the luminosity density and its logarithmic slope, $S$, which involve one and two derivatives of the surface brightness, respectively. Clearly, the optimal smoothing for the surface brightness is not the optimal smoothing for $\nu$ or $S$. Since derivatives are inherently noisier, a larger smoothing parameter is necessary. We are unaware of standard techniques for estimating the optimal smoothing in this situation. However, one is correct in using a subjective choice of the smoothing parameter, as long as the variation in the desired estimate (in this case, $S$) is not unrealistically large and there is no significant bias in the surface brightness spline. Given the uncertainties in estimating the optimal smoothing for $S$, we have tried different smoothings ($\lambda$ in Eqn. 5) which varied by a factor of 10. The conclusions of this paper do not change even over this large range of smoothings. We have chosen a smoothing which is five times the optimal surface brightness smoothing, which corresponds to using 0.022 (in log $\nu$) as the average uncertainty in the "predicted mean error" method.

Once this spline fit is obtained, the derivative of the surface brightness is computed by differentiating the spline and the density is obtained by performing the integration shown in Eqn. 2. To estimate the density slope $d \log \nu / d \log r$, we fit a second smoothing spline to the luminosity density $\nu(r)$, and differentiate that spline. We do not have reliable *a priori* uncertainties for the luminosity density, and, therefore use GCV instead of the predicted mean error to estimate the optimal smoothing for the luminosity density. This second spline is used only to provide a simple and an accurate measurement of the slope. Since the luminosity density, $\nu(r)$, is already represented by a smooth curve, the spline calculated in GCV passes through the data points with little or no smoothing necessary. An ordinary spline (one that is forced to go through each data point) could be used at this point with no difference in the final result. The measured luminosity density slope is *only* affected by the initial smoothing chosen for the spline of the surface brightness.

The uncertainty in these calculations can be estimated using Monte-Carlo simulations. One thousand realizations are drawn from the surface brightness data points, with each point randomly chosen from a Gaussian distribution with the mean given by the initial measured value (in log surface density) and a standard deviation equal to 0.01 dex (0.025 in magnitudes). For each realization, we calculate the luminosity density and logarithmic slope, and thus obtain distributions at each radial value. Choosing the 5% and 95% data value provides the 90% confidence band. These Monte-Carlo simulations do not include large-scale systematic effects such as poor PSF match (see Paper 1 and Kormendy *et al.* 1995 for a study of this), deconvolution problems, and ellipse fitting problems.

Deprojections for four representative galaxies are shown in Fig. 1. The left column shows the surface brightness profiles plotted as a function of the geometric-mean radius, with the points representing the HST values and the solid line the spline fit. The deprojected luminosity densities are shown in the middle column, and the logarithmic derivatives of the luminosity density are shown in the right column. Dotted lines represent the 90% confidence bands as explained above. Only in $d\log\nu/d\log r$ are the confidence bands large enough to be easily seen in all four galaxies. The results suggests that the densities at $0.1''$ are typically accurate to about 2% ($1\sigma$), and the slopes to $\pm 0.1$ (in units of $d\log\nu/d\log r$).

The above calculations have assumed spherical symmetry, although for most of the galaxies studied this assumption is not valid. The deprojection can be done under only the assumption of axisymmetry, and techniques to do this have been outlined in Palmer (1994) but do not yet exist in full. Axisymmetry is the least constraining assumption possible which will provide a unique solution, although possibly not the true one. Deprojection of the surface brightness of a triaxial system requires assumptions about the triaxiality and therefore would not be unique. In any case, we argue below that the results suggest that the galaxies are not triaxial near the center.

We will assume that the galaxies are spheroidal. In this case Eqn. 2 is then valid to an overall multiplicative factor on the RHS of the equation given by $q/p$, where $q$ is the apparent axis ratio, $p$ is the true axis ratio, and $q^2 = cos^2 i + p^2 sin^2 i$ where $i$ is the inclination (edge-on is $90°$). The average ellipticity for this sample is 0.2 ($q = 0.8$), and if we assume an average inclination of $60°$, then the average correction term would be 10%, but will be higher for galaxies which are face-on. Since it is difficult to estimate the inclinations, we will make no such estimate, and therefore will be underestimating the luminosity density by an average of 10%. So long as the assumption of spheroidal geometry is valid, there is no inclination correction for the logarithmic slope.

## 4. Results

Table 1 gives the assumed distance from Faber *et al.* (1996) (col.2), absolute $V$ magnitude (col.3), surface density at $0.1''$ (col.4), radius of maximum curvature (col.5), luminosity density and uncertainty at $0.1''$ (col.6), and logarithmic slope of the luminosity density and uncertainty at $0.1''$ (col.7) for each galaxy in our sample. We do not report results for galaxies which have significant dust contamination. The galaxies marked with a "1" are those which had some dust contamination which we were able to mask out in calculating the surface brightness profile. Since the data come from the uncorrected HST, we trust only the surface brightness data at radii at and beyond $0.1''$.

Fig. 2 plots the distribution function for the slopes at $0.1''$ (solid line). The area under the solid lines is unity. The distribution function estimate is determined from an adaptive kernel density estimate (Silverman 1986). The adaptive kernel provides a non-parametric density estimate on a grid. An initial density estimate is obtained, and then the window width at each grid element is adjusted according to the initial density such that low density regions have a large window width. We use the Epanechnikov kernel (an inverted parabola) for the density estimate.

The result of an adaptive kernel estimate can be highly dependent on the initial smoothing, especially for small samples like ours with $N < 100$. We therefore use least-squares cross-validation to choose the initial smoothing (see Silverman 1986, p. 48). Least-squares cross-validation is similar to GCV (used in the luminosity density estimate above), although here it is applied to density estimation as opposed to the regression problem for which GCV is designed. The cross-validation technique minimizes an estimate of the integrated square error, given by $\int (\tilde{f} - f)^2$, where $\tilde{f}$ is the density estimated from the data, which is dependent on the window width (the smoothing parameter), and $f$ is the underlying function which we are trying to estimate. As in GCV, one point is removed from the sample, and the density is estimated for a given smoothing from the $N - 1$ data points. Repeating this $N$ times and using the average provides an estimate of the underlying function $f$. The smoothing which minimizes the estimated integrated square error is the optimal smoothing. For the data in Fig. 2, the cross-validation estimate of the initial window width is 0.5 in units of logarithmic slope.

The dashed lines are the 90% confidence bands obtained through a bootstrap estimate. The bootstrap technique draws with replacement from the original sample of 42 slopes. Therefore, one bootstrap realization contains 42 points which are not necessarily unique. For each realization we estimate the distribution function and repeat this procedure 500 times. At each grid element in $S$ we then have a distribution of points, from which we choose the 5% and the 95% values which represent the 90% confidence band. These confidence bands reflect only the variance, and not the bias, of the estimate. The bootstrap procedure includes the measurement uncertainties implicitly because they are reflected in the original estimate of the slope. Since the surface brightness data are well measured, the resulting scatter from their uncertainties has little effect on Fig. 2, as the intrinsic range of central slopes is much greater. From the Monte-Carlo simulations in Sec. 3, it is known that the slopes have an uncertainty of $\sim 0.1$. This is much smaller than the intrinsic range of slopes, and we conclude that the 90% confidence bands are a good representation for the true distribution.

To check whether the innermost slope is affected by noise at $0.1''$, we have also estimated an average slope from $0.1$–$0.5''$. The resulting plot is very similar to Fig. 2. In addition, since the sample contains galaxies which vary in distance from 1–300 Mpc, we must check whether the shape of the slope distribution is a reflection of the distribution of galaxy distances. The slopes are then estimated at new radii which correspond to the same physical distance from the center. We use only galaxies closer than 40 Mpc, and at this distance our smallest radius ($0.1''$) corresponds to 19 pc. The distribution of slopes at 19 pc is practically identical to that of Fig. 2, implying that Fig. 2 reflects an actual physical distribution and not an observational effect.

## 5. Discussion

The sample in Paper 1 is divided into two groups of early-type galaxies based on their distribution of surface density slopes and correlations with other physical properties (shown in Faber *et al.* 1996). These two groups are referred to as "core" and "power-law" galaxies. Fig. 2, based on yet more data, also suggests that there are two groups of slopes, consistent with the Paper 1 result. One peak is at $S \equiv -0.8$, and the other is at $-1.9$. Using the DIP test (Hartigan & Hartigan 1985) to check for multimodality, we find that the slope distribution is inconsistent with a unimodal distribution at the 90% confidence level. The DIP test measures the maximum distance between the empirical distribution function and the best fitting unimodal distribution, and the significance level is based on simulations. Whether this bimodality is due to the sample selection or is a true feature of the elliptical galaxies remains to be seen with a larger sample. However, it appears that $v/\sigma$, luminosity, and boxiness are also correlated with $S$, suggesting that the distribution of slopes has a physical basis (Faber *et al.* 1996).

There are two galaxies in the sample of 42 which permit $S = 0$ within their 90% confidence band. Fig. 2 also allows a small number of galaxies with zero central density slopes, but this is mainly due to the smoothing scale used there.

To further study the shape of the luminosity density, Fig. 3 plots the scaled luminosity density fits for all 42 galaxies derived in Sec. 3. Each galaxy is scaled to the radius and luminosity at the radius of maximum curvature in the surface brightness spline. This radius is generally close to $1''$. The two conclusions from the slope distribution in Fig. 2 are also apparent in this figure. First, nearly all galaxies in our sample have a power-law cusp in the luminosity density. Second, the division of slopes into two groups is obvious. There is no such division at radii beyond the radius of maximum curvature (as has been known from ground-based work, since most early-type galaxies are well represented by the de Vaucouleurs profile).

Fig. 4 plots the logarithmic slope of the surface brightness versus the logarithmic slope of the luminosity density, both measured at $0.1''$. The solid line is $d\log\Sigma/d\log r + 1$. It is apparent that the projected central slope cannot be used to infer the true central slope of the luminosity density by assuming, for example, the power-law rule $S = d\log\Sigma/d\log r + 1$. For values of the surface brightness slope greater than 0.8, the actual 3-d slope can be obtained by simply adding unity. However, there is no simple relation for surface brightness slope less than 0.8. This should be obvious since if the luminosity density has logarithmic slope less than 1.0, then the surface brightness will be continuously curving and, therefore, it is not possible to define a unique value for its slope. A similar result was found by Kormendy *et al.* (1995). The conclusion is that it is prudent to use the actual 3-d slope to study galaxy properties instead of the surface brightness slope since there is not a one-to-one correspondence between the two.

Paper 1 and Byun *et al.* (1996) used a parametric form to represent the surface brightness (Eqn. 3). Parametric fitting has the advantage of being able to simply represent the data. For example, in the case of Eqn. 3, only five numbers are needed to represent each galaxy, whereas with the smoothing spline fits one needs the positions of the spline knots and the second derivatives at each data point. However, in parametric fitting one has to beware of possible biases in the specific form used. We have checked the quality of the fit to Eqn. 3 by comparing the parametric and non-parametric profiles. In most cases the parametric and the non-parametric fit to the surface brightness are very similar, suggesting that the parametric fit provides an adequate representation for the surface brightness. However, more importantly we need to estimate the luminosity density, which involves a deconvolution where small deviations from the actual surface brightness can have significant effects.

Since there exists an analytic form for the surface brightness, Eqn. 2 is easily integrated to give the corresponding luminosity density, which can be compared with the non-parametric estimate of $\nu$. Fig. 5 plots the ratio of the two luminosity densities as a function of $d\log\Sigma/d\log r$ at $0.1''$. For values $< 0.2$, the two density estimates have little bias and an rms scatter of around 8%. As the surface brightness slope increases to steeper values, the scatter remains about the same, but the bias increases, with the parametric fit having a lower density by about 10%, and there are a few galaxies with large differences. The discrepancy is most likely due to use of a parametric form for the surface brightness which is slightly incorrect, with the discrepancy amplified during deprojection. Many of the galaxies in our sample contain stellar nuclei and those regions were excluded in the parametric fit of Byun *et al.*. This will lead to a bias in the stellar surface brightness estimate at those radii since part of the stellar light has already been ignored and, therefore, it is not surprising that the galaxies with the largest discrepencies are those with nuclei. In addition, the parametric fitting included a constraint on the light in the central $0.1''$. This constraint assumes that the specific form of the model is the true surface brightness profile into $R = 0$. If the parametric form is not an adequate representation for the galaxy inside the central $0.1''$ then the result may be biased, and in an unknown way since we are not able to obtain the surface brightness profile in this region.

The three galaxies in Fig. 5 which lie well above the solid line have stellar nuclei which had to be excluded in the parametric fitting and, thus, the parametric form does not adequately represent their luminosity density. The discrepancy of NGC 1600 is due to the central light constraint of the parametric fitting biasing the surface brightness at $0.1''$. Fig. 5 plotted at $0.2''$ instead of $0.1''$ shows no bias or large discrepancies, suggesting that the parametric form may be systematically off within the central $0.2''$. This illustrates one of the main virtues of the non-parametric estimate, since the spline smoothing fit to the surface brightness has negligible bias. There are individual galaxies which are poorly represented by the parametric form, but for most of the galaxies, the parametric and the non-parametric fitting give similar surface brightness profiles. Thus, the main conclusions of this paper and Paper I are unaffected by which fitting technique is used.

Fig. 6 plots the non-parametric logarithmic density slope $S(0.1'')$ verses the absolute magnitude. Dimmer galaxies tend to have steeper central cusps. This trend is noted in Faber *et al.* (1996) based on parametric fits to surface brightness slopes, and persists here using non-parametric fits to the density slope. The bimodal slope distribution of Fig. 2 divides galaxies in Fig. 6 into two classes, bright and faint. Faber *et al.* note that these two classes are the classic boxy/non-rotating and disky/rotating classes of Bender *et al.* (1992).

Merritt and Fridman (1996) recently studied the regularity of orbits in triaxial mass distributions with cusps. They show that in profiles with $\rho \propto r^{-2}$ the phase space of orbits is dominated by stochastic orbits, and there are insufficient regular box orbits (flattened in the same direction as the mass distribution) to match the triaxial mass distribution. Since the mass distribution of the stochastic orbits are rounder than the mass distribution of the triaxial galaxy, an equilibrium configuration could not be found. Using these results as a guide, it seems very likely that the steep profile (low luminosity) elliptical galaxies in our sample are not triaxial anywhere near the center (although, since many are known to rotate they may well be oblate). Further evidence for the lack of triaxiality in these galaxies comes from observations at larger radii. Low luminosity galaxies exhibit little minor axis rotation and rotate rapidly (Davies *et al.* 1983, Kormendy & Bender 1995), both of which do not support triaxial structure.

In the case of the shallow profile (and high luminosity) galaxies, the situation is more complex. With milder density cusps with $\rho \propto r^{-1}$, Merritt & Fridman show that orbits are still mainly chaotic, but they mimic regular behavior for a long time before diffusing in phase space. In this case, quasi-equilibrium configurations could be found which may last a substantial fraction of a Hubble time. Even though ellipticals and bulges in our sample are slightly more shallow then the $\rho \propto r^{-1}$ case investigated by Merritt and Fridman, they do not approach the analytic case that would correspond to a Stäckel potential where regular box orbits are known to exist. The presence of a massive dark object would further steepen the potential and accelerate the stochastic diffusion of orbits in such a system (Merritt 1996). There appears to be two possibilities in this case. The galaxies may be evolving slowly under the influence of orbit diffusion and may still be triaxial. Or they may be axisymmetric near the center. However at large radii, high luminosity galaxies exhibit minor axis rotation and rotate slowly (Bertola & Capaccioli 1975, Kormendy & Bender 1995), suggesting that these galaxies may be triaxial at large radii.

While Merritt & Fridman's work only investigated two density profiles and one particular set of axial ratios, their results seem to be generic. However the radial extent and timescales for different mass models for which stochastic orbits will influence the mass distribution of the galaxy has just begun to be explored through the work of Merritt & Fridman, and future work needs to concentrate on cuspy galaxy models.

These results superficially contradict recent work by Tremblay & Merritt (1995) and Ryden (1996) which determined the intrinsic shapes of elliptical galaxies using the projected axis ratio distribution. Both studies find that elliptical galaxies are not consistent with being solely axisymmetric (at significance level of $> 99\%$), and the results are fully consistent with the galaxies being triaxial. The authors did not consider a mixture of axisymmetric and triaxial shapes. These results are due to an absence of nearly round galaxies; it might be possible that different methods for estimating the projected axis ratios would give a different result. This apparent discrepancy may simply reflect the fact that we have focussed on the galaxy centers and the other authors have examined the main bodies of the galaxies. However, Tremblay & Merritt (1996) have divided the sample into two groups based on brightness and find that the faint galaxies are consistent with oblate symmetry whereas the bright galaxies are not, which is further evidence for two families of elliptical galaxies.

In any case, it seems very unlikely that experience gained from the analysis of orbits in static Stäckel potentials or of triaxial objects with analytic cores has much connection to the central regions of real galaxies.

The main conclusions of this paper are:

• Nearly all early-type galaxies looked at with HST have power-law cusps into the innermost measured value.

• There are appears to be a division of early-type galaxies based on their value of the central slope of the luminosity density.

• Many and maybe most elliptical galaxies may be either nearly axisymmetric or spherical near the center, or may slowly evolve due to the influence of stochastic orbits.

We have benefited from discussions with David Merritt and Tema Fridman. We thank David Merritt for his many suggestions about non-parametric techniques and for his comments on previous drafts of this paper.

---

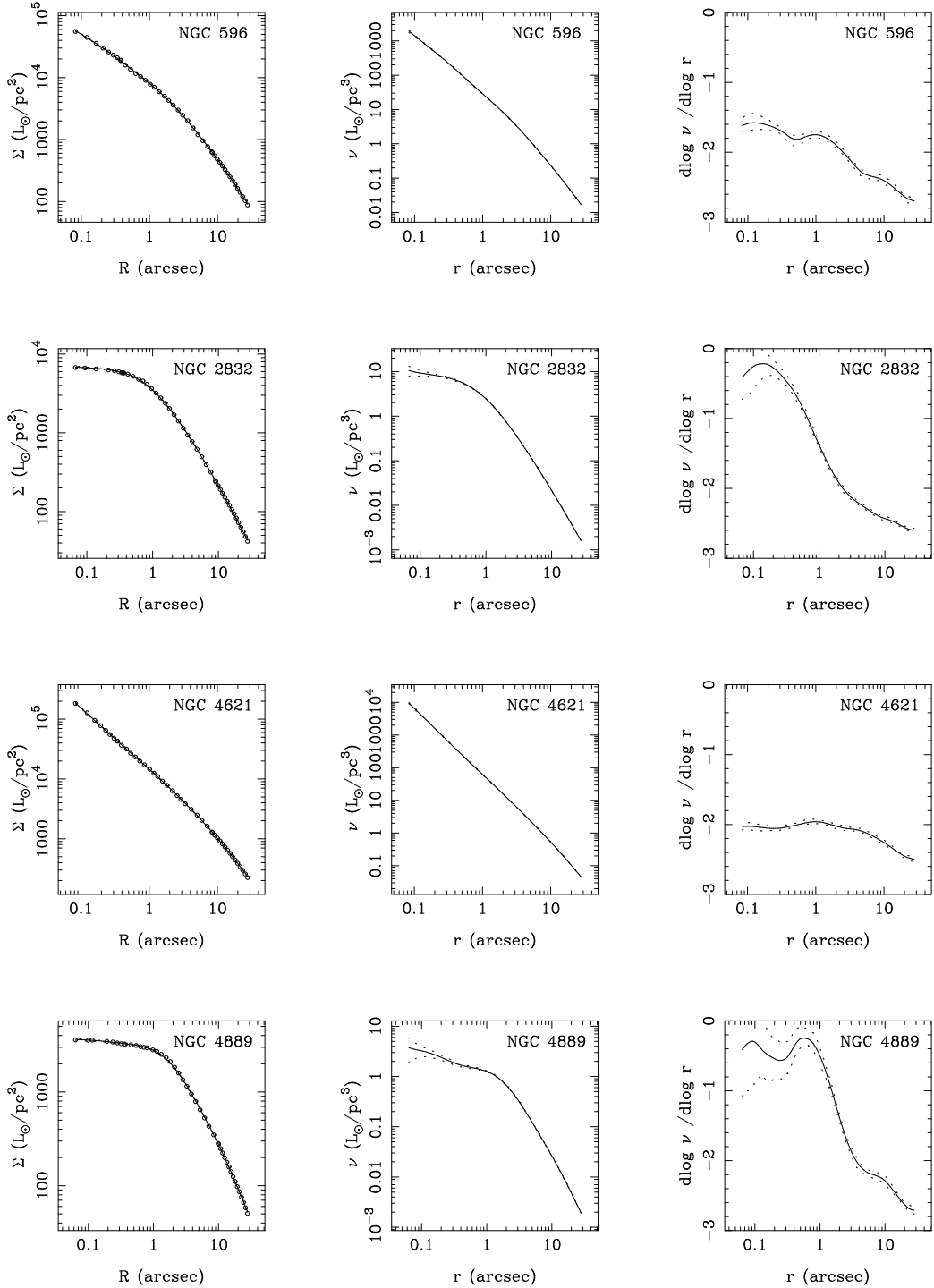

Fig. 1.— Surface brightness, luminosity density, and logarithmic density slope profiles for four of the galaxies in the sample. The points in the left column are the HST surface brightness values from Lauer *et al.* (1995) and the solid lines are the non-parametric estimates. The dotted lines are the 90% confidence bands.

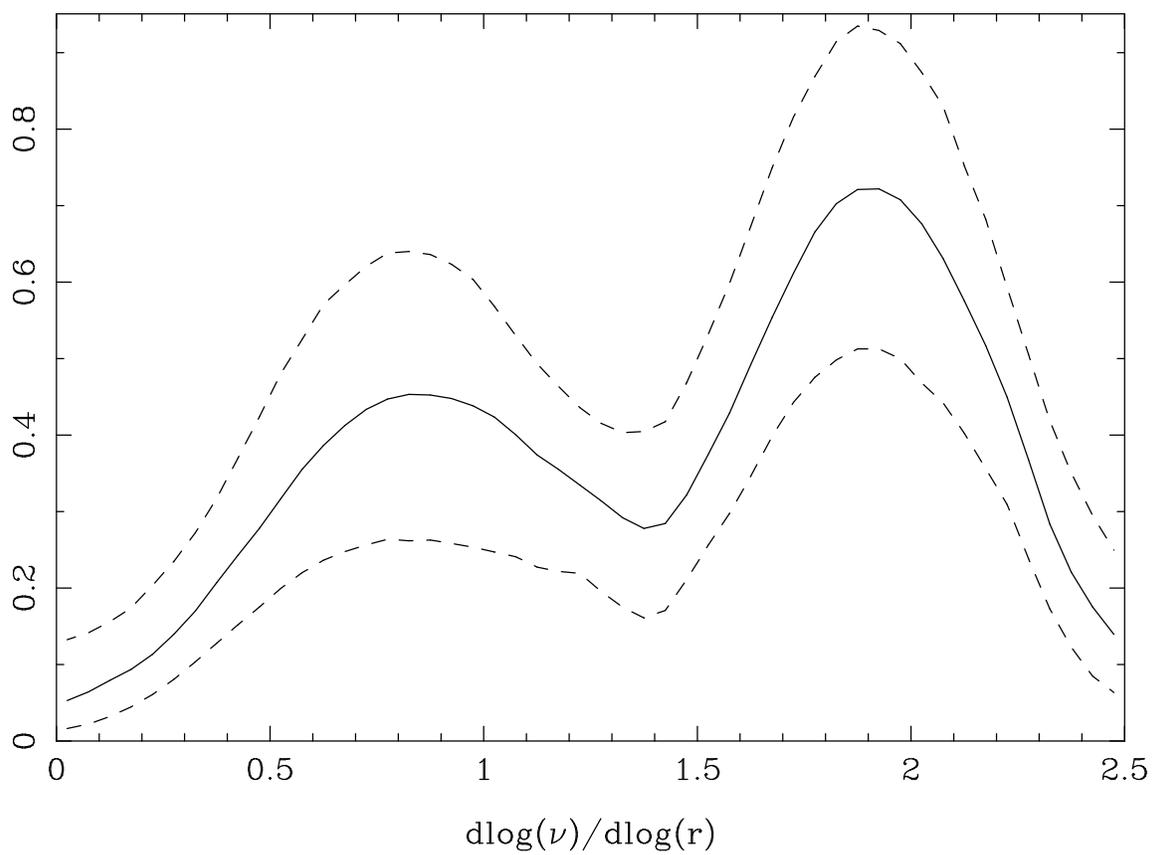

Fig. 2.— Distribution function (solid line) for the logarithmic slope of the luminosity density, measured at $0.1''$. The dashed lines represent the 90% confidence band.

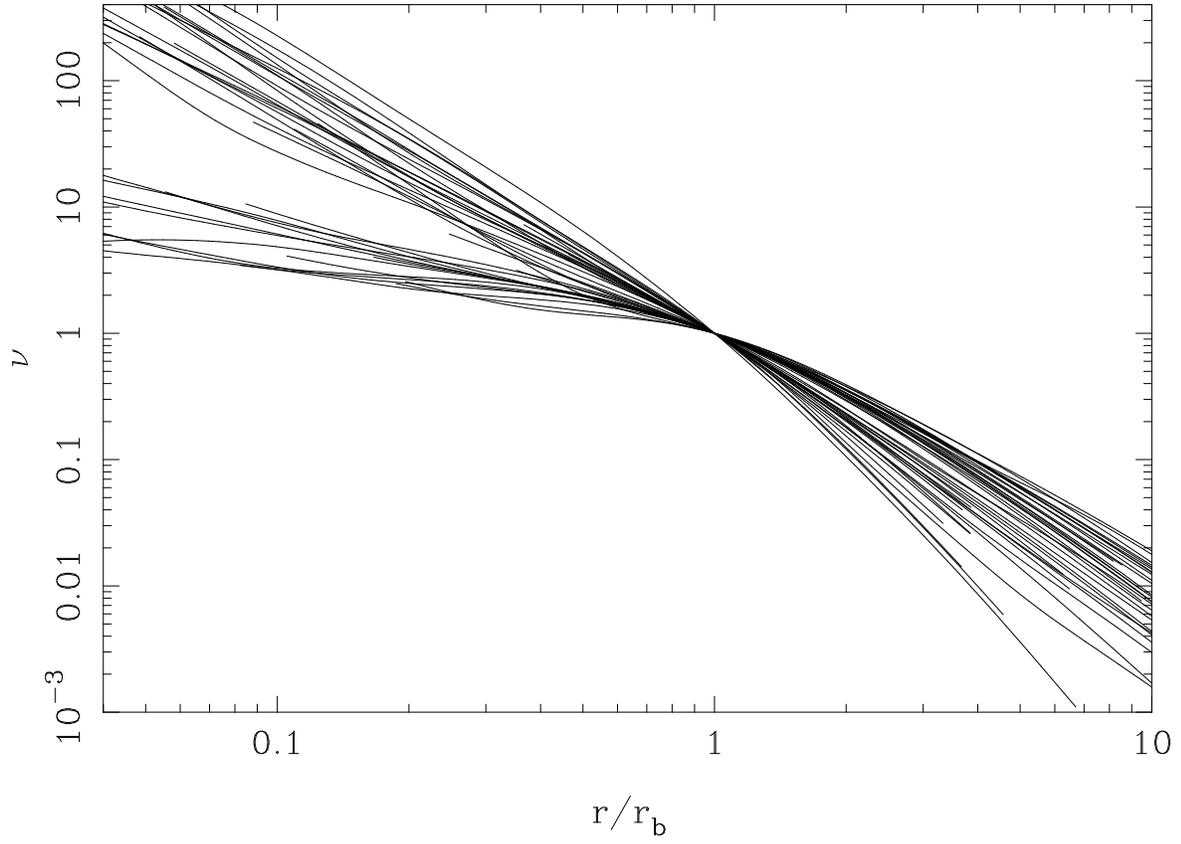

Fig. 3.— Luminosity density profiles of all galaxies in the sample scaled to the radius and the luminosity at the radius of maximum curvature in the surface brightness

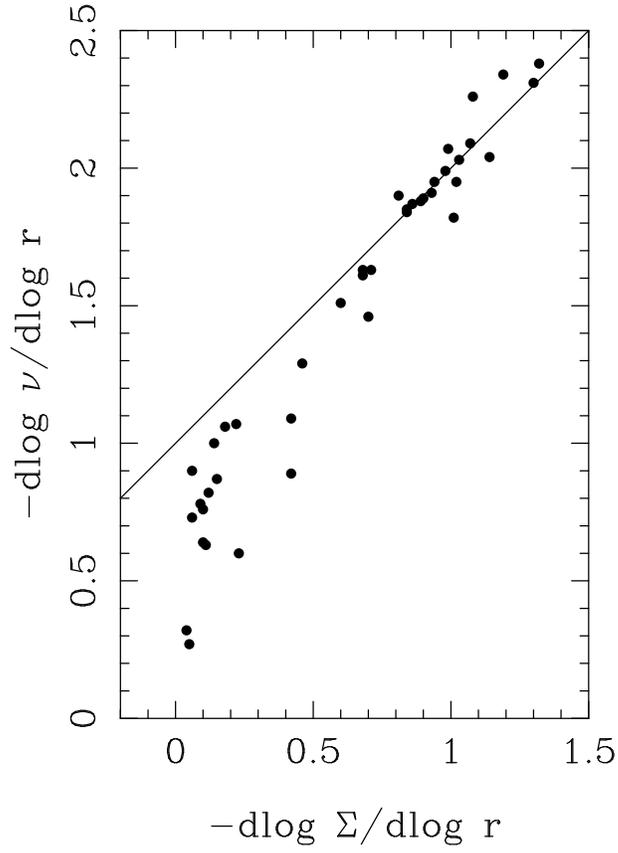

Fig. 4.— Non-parametric estimate of the luminosity density slope verses the surface density slope at $0.1''$. The solid-line is $1 + d\log\Sigma/d\log r$.

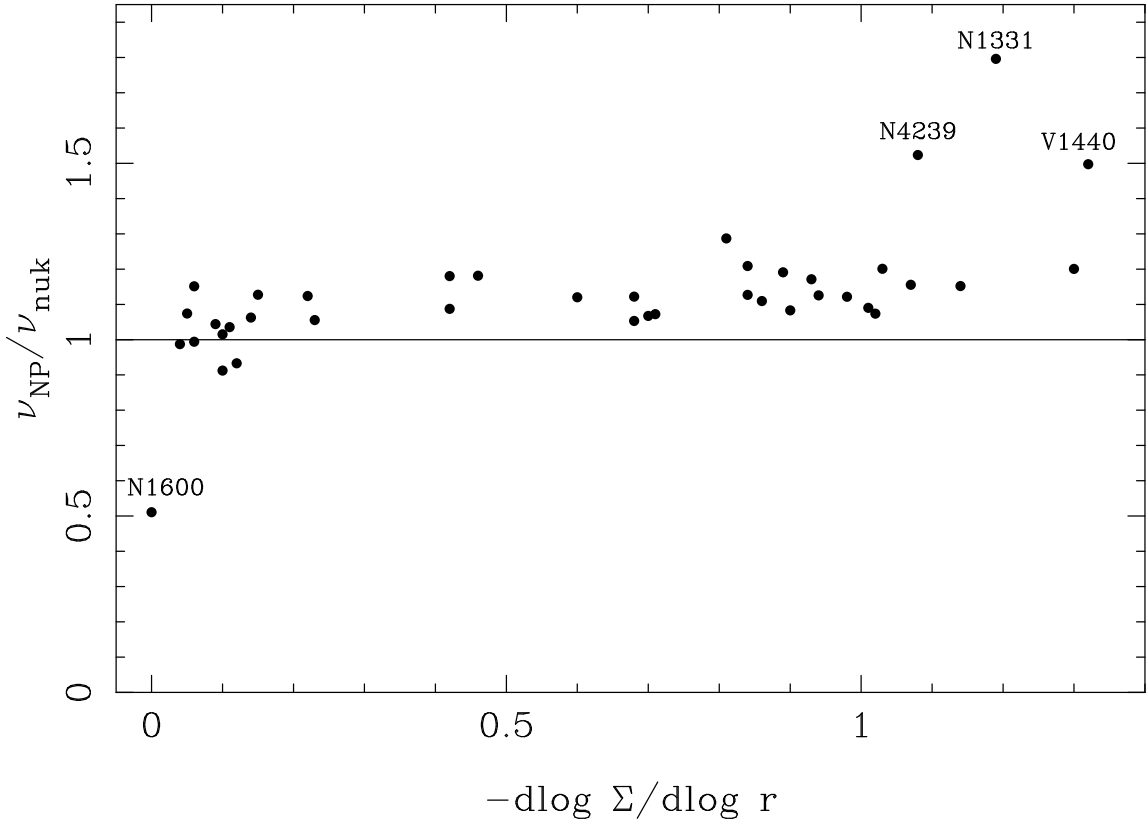

Fig. 5.— Ratio of the luminosity densities measured non-parametrically and from a parametric fit as a function of $d\log\Sigma/d\log r$.

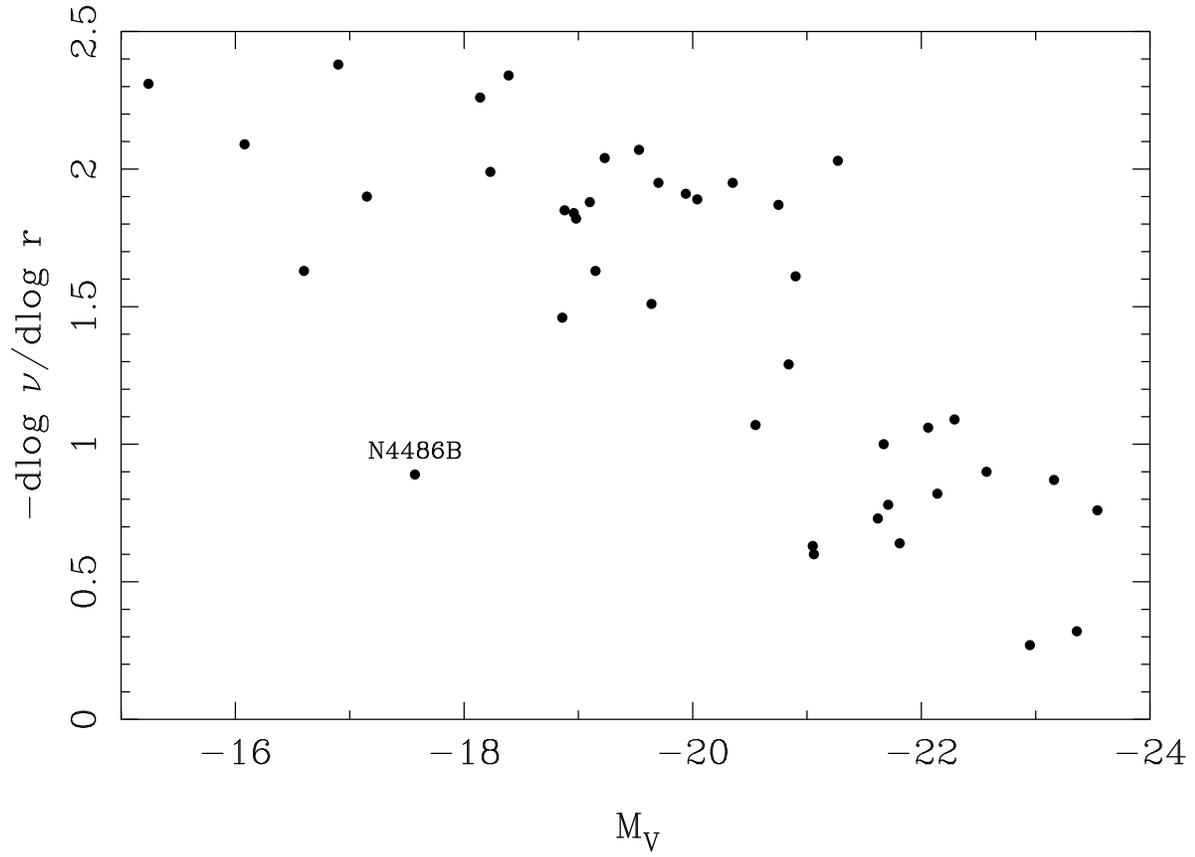

Fig. 6.— Luminosity density slope at 0.1″ measured non-parametrically verses absolute magnitude.

TABLE 1
Galaxy Parameters

| Galaxy | Dist Mpc | $M_V$ | $\Sigma(r=0.1'')$ $L_\odot/\mathrm{pc}^2$ | $r_b$ ('') | $\nu(r=0.1'')$ $L_\odot/\mathrm{pc}^3$ | $-\frac{d\log\nu}{d\log r}$ $r=0.1''$ |
|---|---|---|---|---|---|---|
| Abell 1020 | 244.0 | −22.29 | $7.9\times10^3$ | 0.25 | $1.4(\pm0.07)\times10^1$ | $1.10\pm0.11$ |
| Abell 1831 | 281.0 | −23.16 | $1.8\times10^3$ | 0.44 | $1.2(\pm0.10)\times10^0$ | $0.86\pm0.17$ |
| Abell 2052 | 133.0 | −22.66 | ... | ... | ... | ... |
| M 32 | 0.8 | −16.60 | $7.1\times10^5$ | 0.48 | $4.9(\pm0.13)\times10^5$ | $1.63\pm0.04$ |
| NGC 224 | 0.8 | −19.82 | ... | ... | ... | ... |
| NGC 524 | 29.0 | −21.51 | ... | ... | ... | ... |
| NGC 596 | 18.0 | −20.90 | $5.1\times10^4$ | 2.64 | $1.5(\pm0.05)\times10^3$ | $1.62\pm0.05$ |
| NGC 720 | 25.0 | −21.62 | $7.3\times10^3$ | 2.90 | $2.2(\pm0.39)\times10^1$ | $0.73\pm0.32$ |
| NGC 1023 | 10.7 | −20.14 | ... | ... | ... | ... |
| NGC 1052 | 18.0 | −19.88 | ... | ... | ... | ... |
| NGC 1172 | 30.3 | −20.74 | ... | ... | ... | ... |
| NGC 1331 | 22.0 | −18.39 | $6.3\times10^3$ | 4.32 | $2.0(\pm0.06)\times10^2$ | $2.34\pm0.05$ |
| NGC 1399 | 17.0 | −21.71 | $1.5\times10^4$ | 2.68 | $9.5(\pm1.04)\times10^1$ | $0.78\pm0.14$ |
| NGC 1400 | 24.0 | −21.06 | $3.8\times10^4$ | 0.30 | $4.7(\pm0.17)\times10^2$ | $0.60\pm0.08$ |
| NGC 1426 | 22.0 | −20.35 | $4.4\times10^4$ | 1.35 | $1.3(\pm0.03)\times10^3$ | $1.95\pm0.03$ |
| NGC 1600 | 50.7 | −22.70 | $3.2\times10^3$ | 2.92 | $3.3(\pm0.78)\times10^0$ | $-0.19\pm0.53$ |
| NGC 1700 | 38.0 | −21.65 | ... | ... | ... | ... |
| NGC 2636 | 31.0 | −18.86 | $1.8\times10^4$ | 6.08 | $3.2(\pm0.15)\times10^2$ | $1.46\pm0.17$ |
| NGC 2832 | 86.0 | −22.95 | $6.7\times10^3$ | 0.82 | $9.5(\pm0.87)\times10^0$ | $0.27\pm0.17$ |
| NGC 2841 | 14.0 | −19.86 | ... | ... | ... | ... |
| NGC 3115 | 8.9 | −20.75 | $2.3\times10^5$ | 5.63 | $1.6(\pm0.05)\times10^4$ | $1.87\pm0.03$ |
| NGC 3377 | 10.0 | −19.70 | $2.3\times10^5$ | 0.21 | $1.6(\pm0.04)\times10^4$ | $1.95\pm0.05$ |
| NGC 3379 | 10.4 | −20.55 | $3.8\times10^4$ | 0.98 | $8.1(\pm0.48)\times10^2$ | $1.07\pm0.06$ |
| NGC 3384[1] | 10.4 | −19.53 | $1.1\times10^5$ | 1.66 | $6.9(\pm0.18)\times10^3$ | $2.07\pm0.04$ |
| NGC 3599 | 20.8 | −19.71 | ... | ... | ... | ... |
| NGC 3605 | 24.0 | −19.15 | $2.1\times10^4$ | 0.89 | $4.5(\pm0.13)\times10^2$ | $1.63\pm0.06$ |
| NGC 3608[1] | 24.0 | −20.84 | $4.2\times10^4$ | 0.36 | $7.1(\pm0.33)\times10^2$ | $1.29\pm0.05$ |
| NGC 4168 | 36.9 | −21.76 | ... | ... | ... | ... |
| NGC 4150 | 3.0 | −15.15 | ... | ... | ... | ... |
| NGC 4239 | 15.8 | −18.14 | $9.4\times10^3$ | 0.71 | $4.0(\pm0.14)\times10^2$ | $2.26\pm0.07$ |
| NGC 4365[1] | 22.5 | −22.06 | $5.4\times10^4$ | 1.45 | $4.3(\pm0.34)\times10^2$ | $1.06\pm0.09$ |
| NGC 4387 | 15.8 | −18.88 | $1.8\times10^4$ | 4.44 | $6.7(\pm0.18)\times10^2$ | $1.85\pm0.04$ |
| NGC 4434 | 15.8 | −18.96 | $3.7\times10^4$ | 7.30 | $1.3(\pm0.04)\times10^3$ | $1.84\pm0.04$ |
| NGC 4458[1] | 15.8 | −18.98 | $6.4\times10^4$ | 0.17 | $2.8(\pm0.08)\times10^3$ | $1.82\pm0.09$ |
| NGC 4464 | 15.8 | −18.23 | $5.5\times10^4$ | 7.59 | $2.2(\pm0.06)\times10^3$ | $1.99\pm0.06$ |
| NGC 4467 | 15.8 | −17.04 | ... | ... | ... | ... |
| NGC 4472[1] | 15.8 | −22.57 | $1.3\times10^4$ | 2.61 | $6.2(\pm1.03)\times10^1$ | $0.90\pm0.19$ |
| NGC 4473 | 15.8 | −22.57 | ... | ... | ... | ... |
| NGC 4478 | 15.8 | −19.64 | $3.3\times10^4$ | 7.23 | $9.8(\pm0.37)\times10^2$ | $1.51\pm0.06$ |
| NGC 4486 | 15.8 | −22.38 | ... | ... | ... | ... |
| NGC 4486B | 15.8 | −17.57 | $5.1\times10^4$ | 4.19 | $1.4(\pm0.04)\times10^3$ | $0.89\pm0.09$ |
| NGC 4550 | 15.8 | −17.57 | ... | ... | ... | ... |
| NGC 4551 | 15.8 | −19.10 | $2.2\times10^4$ | 7.56 | $8.6(\pm0.22)\times10^2$ | $1.88\pm0.04$ |
| NGC 4552[1] | 15.8 | −21.05 | $4.0\times10^4$ | 0.46 | $4.2(\pm0.32)\times10^2$ | $0.63\pm0.14$ |
| NGC 4564[1] | 15.8 | −19.94 | $8.5\times10^4$ | 6.39 | $3.3(\pm0.09)\times10^3$ | $1.91\pm0.03$ |
| NGC 4570[1] | 15.8 | −20.04 | $8.7\times10^4$ | 2.60 | $3.3(\pm0.08)\times10^3$ | $1.89\pm0.04$ |
| NGC 4594 | 9.7 | −21.78 | ... | ... | ... | ... |
| NGC 4621[1] | 15.8 | −21.27 | $1.6\times10^5$ | 7.20 | $6.7(\pm0.14)\times10^3$ | $2.03\pm0.03$ |
| NGC 4636 | 15.8 | −21.67 | $7.7\times10^3$ | 3.29 | $7.5(\pm0.62)\times10^1$ | $1.00\pm0.09$ |
| NGC 4649 | 15.8 | −22.14 | $1.2\times10^4$ | 3.41 | $1.1(\pm0.12)\times10^2$ | $0.82\pm0.16$ |
| NGC 4697 | 11.0 | −21.02 | ... | ... | ... | ... |
| NGC 4742[1] | 13.0 | −19.23 | $1.4\times10^5$ | 1.43 | $8.0(\pm0.23)\times10^3$ | $2.04\pm0.10$ |
| NGC 4826 | 6.0 | −20.31 | ... | ... | ... | ... |
| NGC 4874 | 90.0 | −23.54 | $1.7\times10^3$ | 2.68 | $2.3(\pm0.18)\times10^0$ | $0.76\pm0.17$ |
| NGC 4889 | 90.0 | −23.36 | $3.6\times10^3$ | 1.76 | $3.2(\pm0.43)\times10^0$ | $0.33\pm0.44$ |
| NGC 5322 | 20.0 | −21.32 | ... | ... | ... | ... |
| NGC 5813[1] | 28.0 | −21.81 | $1.6\times10^4$ | 0.75 | $8.6(\pm0.77)\times10^1$ | $0.64\pm0.15$ |
| NGC 5845 | 28.0 | −19.87 | ... | ... | ... | ... |
| NGC 6166 | 113.0 | −23.47 | ... | ... | ... | ... |
| NGC 7332 | 15.8 | −19.91 | ... | ... | ... | ... |
| NGC 7457 | 14.1 | −18.57 | ... | ... | ... | ... |
| NGC 7768 | 103.6 | −22.93 | ... | ... | ... | ... |
| VCC 1199[1] | 15.8 | −15.24 | $1.2\times10^4$ | 1.59 | $5.9(\pm0.15)\times10^2$ | $2.31\pm0.05$ |
| VCC 1440 | 15.8 | −16.90 | $1.1\times10^4$ | 2.01 | $5.4(\pm0.17)\times10^2$ | $2.38\pm0.05$ |
| VCC 1545 | 15.8 | −17.15 | $2.5\times10^3$ | 0.78 | $8.5(\pm0.34)\times10^1$ | $1.90\pm0.06$ |
| VCC 1627 | 15.8 | −16.08 | $1.3\times10^4$ | 2.68 | $5.7(\pm0.19)\times10^2$ | $2.08\pm0.04$ |